\documentclass[aps,prl,amsmath,amssymb,reprint,superscriptaddress,floatfix]{revtex4-2}
\AtBeginDocument{\def\selectlanguage#1{}}%
\usepackage{graphicx}%
\usepackage{overpic}%
\usepackage{array}[=2016-10-06]%
\usepackage{dcolumn}%
\usepackage{bm}%
\usepackage{hyperref}%
\usepackage{xcolor}
\usepackage[normalem]{ulem}

\definecolor{amethyst}{rgb}{0,0,0}
\definecolor{turquoise}{rgb}{0,0,0}
\newcommand{\DLN}[1]{\textcolor{amethyst}{#1}}
\newcommand{\ZY}[1]{\textcolor{turquoise}{#1}}

\begin{document}

	\preprint{APS/123-QED}

   	\title{
Turbulence Decay Intensifies Clustering of Bubbles and Particles}

	\author{Dante Lamenza Naylor}
 	
         \affiliation{%
		Department of Physics, University of Massachusetts Amherst, MA 01003, USA
	}%
    \affiliation{%
		Department of Physics, Brown University, Providence, RI 02912, USA.
	}%
   \author{Zhiyu Yang}
         \affiliation{%
		Department of Physics, University of Massachusetts Amherst, MA 01003, USA
	}%
    \author{Lei Yi}
         \affiliation{%
		Department of Physics, University of Massachusetts Amherst, MA 01003, USA
	}%
        \author{Rodolfo Ostilla-M\'onico}
        \affiliation{Escuela Superior de Ingeniería, Universidad de Cádiz, Cádiz, Spain.}
  \author{Enrico Calzavarini}
        \affiliation{Université de Lille, ULR 7512–Unité de Mécanique de Lille Joseph Boussinesq (UML), F-59000 Lille, France}
	\author{Varghese Mathai} \thanks{vmathai@umass.edu}
		\affiliation{%
		Department of Physics, University of Massachusetts Amherst, MA 01003, USA
	}%
	\affiliation{%
		Department of Mechanical Engineering,  University of Massachusetts Amherst, MA 01003, USA.}%
	
	\date{\today}
	
	\begin{abstract}
\noindent {

\textcolor{black}{Our understanding of inertial particle dynamics in turbulence is mostly based on flows held in a statistically stationary state, a particular regime that differs from many natural flows where energy input can {often be} intermittent or cyclic, or may abruptly cease. Here we investigate inertial particle and bubble dynamics in freely decaying turbulence through complementary experiments and direct numerical simulations. While particle accelerations decay monotonically in time, we find evidence that the clustering can exhibit a non-monotonic evolution, intensifying sharply before subsequently weakening. We demonstrate that both the acceleration and clustering behaviors can be mapped onto their counterparts in statistically stationary turbulence using a dynamic rescaling of the evolving length and time scales of the turbulence.
Validity conditions for the dynamic rescaling, satisfied by both the experimental and numerical datasets, are derived. The proposed mappings remain applicable across a broad range of density ratios, from light to heavy particles, and particle sizes spanning two orders of magnitude in Stokes number.
}
}

	\end{abstract}
	

	\maketitle
	\newpage

Turbulent flows laden with small solid particles or bubbles occur widely in nature and industry, from microplastics, plankton and gas bubbles in the ocean \cite{brouzet2021laboratory,giurgiu_full_2024,schmitt2008intermittent}, to droplets in atmospheric clouds \cite{bodenschatz2010can,stjern2023turbulent}. 
Particle dynamics from a Lagrangian viewpoint, in particular the accelerations and preferential concentration in homogeneous isotropic turbulence (HIT), have been investigated extensively in simulations and experiments \cite{toschi2009lagrangian,mathai2020BubblyBuoyantParticlea,brandt2022ParticleLadenTurbulenceProgressb,calzavarini_dimensionality_2008,bec2007HeavyParticleConcentration,zaichik2007RefinementProbabilityDensity,bragg2015mechanisms}.
 Non-neutrally buoyant, finite-inertia particles deviate from tracer-like behavior and preferentially concentrate, even though the advecting flow is itself incompressible \cite{bec_acceleration_2006, volk2008LaserDopplerMeasurement,tagawa_three-dimensional_2012,mathai_microbubbles_2016}. For instance, heavy particles preferentially sample {high-deformation (hyperbolic)}, low-vorticity {(elliptical)} regions, whereas light particles and bubbles concentrate in high-vorticity, {weakly hyperbolic} regions \cite{eaton1994preferential}. The spatial structure of clusters also differs: bubbles and light particles concentrate in filamentary structures, while heavy particles form sheet-like or wall-like clusters \cite{ayyalasomayajula_lagrangian_2006, ayyalasomayajula_modeling_2008}; %
 such features are commonly quantified by the correlation dimension, $D_2$ \cite{calzavarini_dimensionality_2008,bec2003FractalClusteringInertiala}.

Studies of particle-laden turbulent flows have been mostly limited to statistically stationary {turbulence (SST) or forcings that are steady on average \cite{zapata_turbulence_2024,toschi2009lagrangian}}. {Statistical stationarity, however, is an idealization \cite{lesieur_turbulence_2008}: once forcing weakens or stops,  free decay becomes the generic state of turbulence \cite{eyink_free_2000,george_exponential_2009,lohse_crossover_1994,valente_decay_2011,goto2015energy,panickacheril_john_laws_2022,sinhuber2015decay}}.  Experiments probing decaying particle-laden turbulence require the forcing to be shut off within a fraction of the flow evolution time scale.
 Numerical explorations have been conducted on heavy particles in steadily forced, spatially non-uniform turbulent flow, where the spontaneous large-scale unsteadiness was shown to drive extreme clustering states \cite{zapata_turbulence_2024}.
 In a very recent numerical study \cite{obligado2026evolution},  preferential concentration of heavy particles in decaying turbulence was investigated. While the particle slip velocity was found to co-evolve with the instantaneous Stokes number, the clustering statistics based on Vorono\"i tessellations retained a dependence on the flow history. {It thus remains unclear whether a unified mapping can be constructed for particle dynamics in decaying turbulence -- one that encompasses both the accelerations and the clustering, and spans arbitrary particle densities.}

\begin{figure*}[!htbp]
\centering
\includegraphics[width = 0.99 \textwidth]{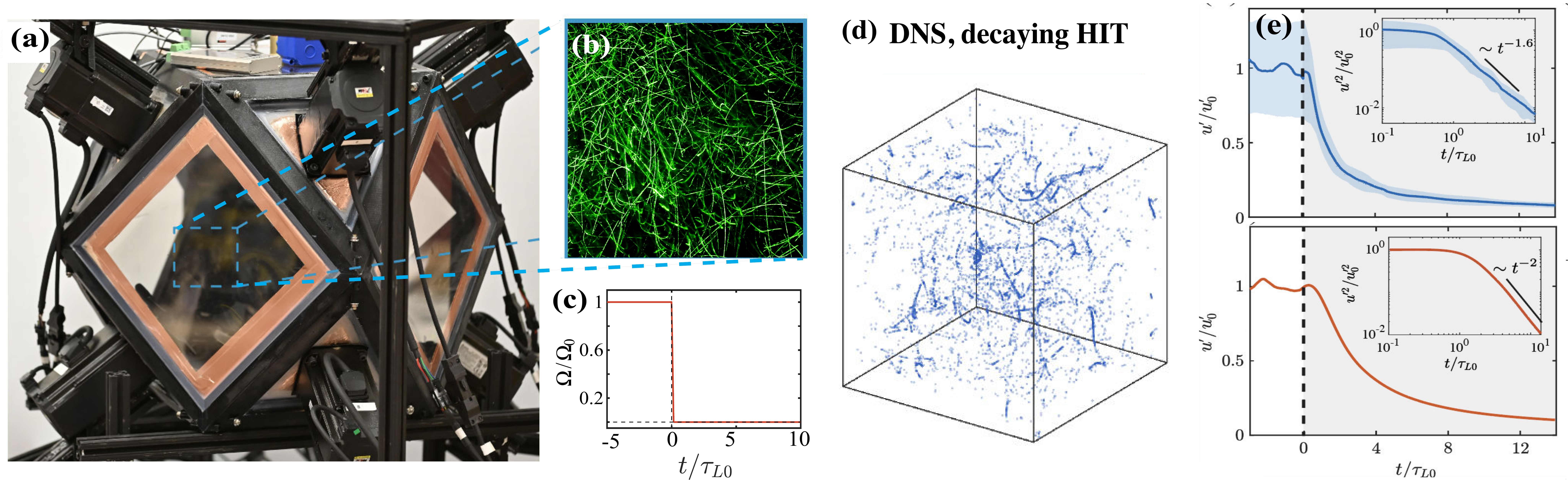}
\caption{Experiments and numerical simulations of particle-laden freely decaying turbulence. (a) Turbulence chamber generating a peak $\text{Re}_{\lambda 0}\approx470$ in the stationary state, and (b) tracer trajectories tracked with a 532\,nm laser sheet. (c) Abrupt cessation of the forcing initiates free decay. (d) Particle clusters from DNS at $\text{Re}_{\lambda 0}=180$ with the same abrupt decay; the simulated flow underlies the Lagrangian particle equations. (e) Velocity decay $u'/u'_0$ vs.\ $t/\tau_{L0}$ ($t=0$ at forcing cessation) for EXP ({\it upper}) and DNS ({\it lower}); insets: ${u'}^2/{u'_0}^2$ on log--log axes, showing power-law decay beyond an initial transient.}
 \label{fig-1}
  \vspace{-.4cm}
 \end{figure*}

In this Letter, we combine experiments and direct numerical simulations of freely decaying isotropic turbulence (FDT) laden with inertial particles and bubbles. We show that  the particle and bubble accelerations decay monotonically with the turbulence, while the preferential concentration can evolve non-monotonically, creating extreme clusters even as the flow greatly weakens. We reveal how a dynamic rescaling framework allows a direct mapping of  both accelerations and clustering from FDT onto their SST counterparts. Validity conditions for the quasi-stationary (QS) assumption are derived, and we {provide hints about} departures from QS at late times and for high-inertia particles.

\noindent
{\it Experiments:} The experimental facility is a cubeoctahedron turbulence chamber (Fig.~\ref{fig-1}a), detailed in End Matter, Appendix~A. The apparatus was specifically designed to obtain nearly isotropic turbulence  in a central volume  with a Taylor-scale Reynolds number, Re$_{\lambda0} \approx 470$, and is capable of abrupt cessation of forcing over a short duration $\delta t \approx 10$ ms that is significantly shorter than the temporal evolution time scale, $\sim \mathcal{O}(1)$s, of the turbulence (Fig.~\ref{fig-1}c).  {The heavy, inertial particles studied here span three mass densities, $\rho_p = 170$, $860$, and $1400$~kg/m$^3$.} {Figure~\ref{fig-1}b gives an example streak-image of the particle trajectories} showing semi-ballistic paths. 
{Once the motors are impulsively brought to a stop, the energy injection to the fluid is halted, allowing the turbulence to freely decay and the particles to evolve with it (Fig.~\ref{fig-1}e).}  
All quantities from the initial statistically stationary state (SSS) will be denoted using subscript $(\cdot)_0$, and subscripts $(\cdot)_f$ and $(\cdot)_p$ will refer to fluid and particle variables, respectively.

 \begin{figure*}[!tbp]
\centering
\includegraphics[width = 0.99 \textwidth]{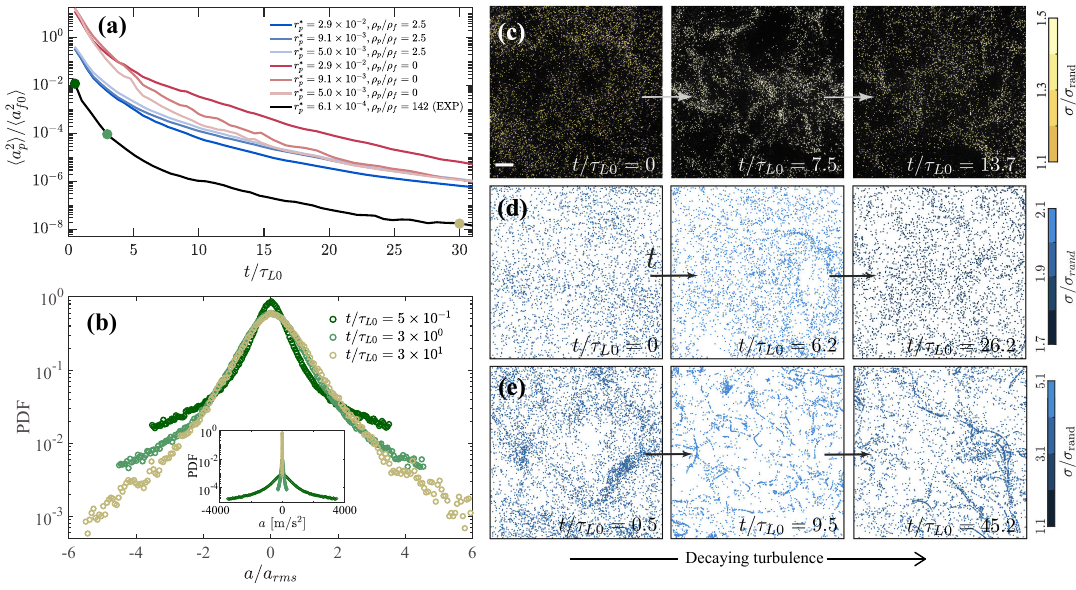}
\caption{Accelerations and clustering in decaying turbulence, from experiments (EXP) and simulations (DNS). (a) Normalized acceleration variance from DNS ($\text{Re}_{\lambda 0}=180$) and EXP ($\text{Re}_{\lambda 0}=470$), for all densities and inertias probed; filled circles mark three decay times. (b) Corresponding horizontal-acceleration {\it pdfs}; inset: in physical units ($\mathrm{m/s^2}$). (c) EXP snapshots at $t/\tau_{L0}=0,\,7.5,\,13.7$ for heavy particles, colored by the Vorono\"i-area standard deviation $\sigma/\sigma_{\mathrm{rand}}$; scale bar $10$~mm. (d,e) DNS snapshots of heavy particles and of bubbles ($\rho_p/\rho_f \to 0$) in a $0.5H\times 0.5H\times H$ sub-volume, with $\sigma/\sigma_{\mathrm{rand}}$ from cell volumes. In all cases clustering first intensifies, then decays.}
 \label{fig:collection}
 \vspace{-.3cm}
 \end{figure*}

\noindent
{\it Numerical Simulations:} We also perform direct numerical simulations (DNS) of the incompressible Navier--Stokes equations \cite{eswaran1988examination,chouippe_forcing_2015} (End Matter, Appendix~B), with a Taylor-Reynolds number Re$_{\lambda}=180$.
Particles of various densities, ranging from bubbles to the very heavy limit, were simulated. A snapshot of the spatial distribution of inertial bubbles in the SSS is shown in Fig.~\ref{fig-1}d.  Once the stationary state was established, the forcing was abruptly removed, allowing the turbulence to {freely decay} (see Sec.~\ref{sec:NumericalDetails} of the Supplemental Material \cite{supp}). {For the particles, we use a model equation of motion considering passively advected spheres acted upon by inertial and viscous
forces \cite{calzavarini_acceleration_2009,mathai_microbubbles_2016}:
\vspace{-0.15cm}
\begin{equation}
(\rho_p + c_a \rho_f) \frac{\mathrm{d}^2  \boldsymbol{x_p}}{\mathrm{d} t^2} =(1 + c_a)\rho_f \frac{\mathrm{D}  \boldsymbol{u}}{\mathrm{D} t} +  \frac{{18} \rho_f \nu }{d_p^2} ( \boldsymbol{u} - \frac{\mathrm{d}  \boldsymbol{x_p}}{\mathrm{d} t}).
\label{eq:StParticles}
\end{equation}
Here, $\boldsymbol{v}(t)\equiv\text{d}\boldsymbol{x}_p(t)/\text{d}t$ is the particle velocity, $\boldsymbol{u}(\boldsymbol{x}_p(t),t)$ is the velocity of the fluid evaluated at the particle position, $\text{D}/\text{D}t$ denotes the temporal fluid material derivative, $c_a=1/2$ is the added mass coefficient, $\rho_p$ and $\rho_f$ are the particle and fluid mass densities, and $d_p=2r_p$ is the particle diameter.

 {Figure \ref{fig-1}e shows time series of velocity fluctuations from the experiments (upper figure) and the DNS (lower figure) as the turbulence decays. {Beyond the initial transient phase, the decay of rms velocity fluctuations} is {well described} by a power law $u'^{2}\sim t^{-n}$. For the DNS, the integral length scale $L$ remains nearly constant ($L\sim t^{m}$, $m\approx 0$), consistent with prior studies of decaying HIT \cite{panickacheril_john_laws_2022}. 
{Consequently, the Kolmogorov length scale $\eta(t)=(\nu^{3}/\varepsilon)^{1/4}$ grows in time. The dissipative-to-integral length scale ratio, $\eta(t)/L$, also grows as the turbulence decays, which we capture using the relation:} 
 \begin{equation}
\frac{\eta(t)}{\eta_{0}}   = \!\left[1 + C_{\eta}\!\left(\frac{t}{\tau_{L0}}\right)^{\!\alpha}\right]^{\!1/2}\!,
  \label{eq:KolmEvol}
\end{equation}
with $C_{\eta}=0.568$ and $\alpha=1.38$, and $\eta_0$ and $\tau_{L0}$ denote the initial Kolmogorov length and the initial large-eddy turnover time of the SST. %

\begin{figure}[!htbp]
\centering
\includegraphics[width=0.47\textwidth]{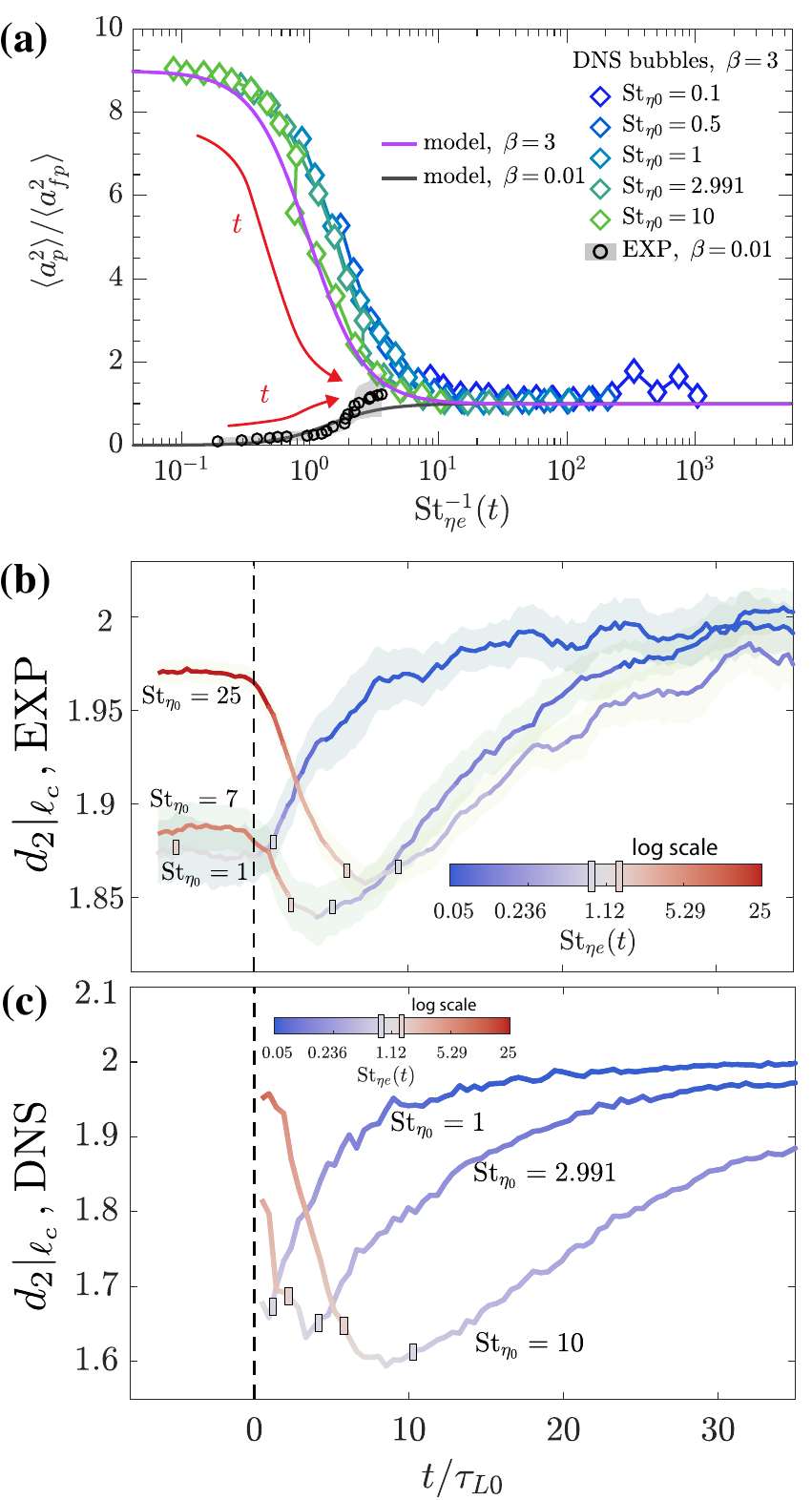}%
\caption{Evolving accelerations and correlation dimension in decaying turbulence. (a) $\langle a_p^2\rangle/\langle a_{fp}^2\rangle$ vs.\ $\text{St}_{\eta e}^{-1}(t)$ for bubbles (DNS, $\beta =3$) and heavy particles (EXP, $\beta=0.01$). Model curves (Eq.~\ref{eq:response_accel}) for bubbles (purple) and heavy particles (dark gray). (b,c) Evolution of correlation dimension, $d_2(t)$ vs. time, extracted from experiment (EXP) and DNS, respectively, and spanning  a range of initial Stokes numbers, $\text{St}_{\eta 0}$. The evolution curves are colored by their respective evolving Stokes numbers, $\text{St}_{\eta e}(t)$. {Vertical bars mark where each curve sweeps across the $\text{St}_{\eta e} \in [0.8,1.2]$ band.}}
 \label{fig:3}
  \vspace{-.1cm}
 \end{figure}
 
{To give a sense of {the} calming of the flow, the centrifugal forcing of the evolving Kolmogorov eddy on a particle, $a_c \sim  \eta(t)/\tau_\eta(t)^2$, has been suppressed by nearly two orders of magnitude after a decay time of $10 \tau_{L0}$. In Fig.~\ref{fig:collection}a, we quantify the time evolution of the particle and bubble acceleration variance. The acceleration variance decreases smoothly and monotonically over time while the small-scale vortex filaments responsible for the extreme accelerations also die out in time. Figure~\ref{fig:collection}b shows the probability density function ({\it pdf}) of the horizontal particle acceleration at three decay times, $t/\tau_{L0} = 0.5$, 3, and 30 (corresponding to the circles in Fig.~\ref{fig:collection}a). The normalized {\it pdfs} show wide, non-Gaussian tails that are progressively suppressed as the turbulence decays. In physical units, the accelerations reach values of up to $400\,g$ (inset of Fig.~\ref{fig:collection}b).

Although the particle and bubble accelerations decay monotonically with the flow, the clustering response can be contrary to this. Clusters not only persist during the decay, they can transiently \emph{intensify} as the carrier turbulence decays, as shown in the time series snapshots (Fig.~\ref{fig:collection}c--e) of the heavy particle and bubble distributions. These are also quantified through the growing standard deviation of the Vorono\"i  volumes of the particle positions $\sigma$, represented by the colorbars in Fig.~\ref{fig:collection}c-e. For heavy particles (experiments), the clustering intensified until $t/\tau_{L0} = 7.5$ as shown by the middle snapshot of the {upper row}. At this instant, the turbulent kinetic energy has fallen to $k \approx 0.01 k_0$. For later times (by $t/\tau_{L0} = 13.7$) we observe that the clustering has decreased. 
A similar response is observed in the DNS for heavy particles (d) and for bubbles (e). These effects are most pronounced for the bubbles, with a clearly visible peak of $\sigma/\sigma_{\mathrm{rand}} = 5.1$ at  $t/\tau_{L0} = 9.5$ before gradually declining to $\sigma/\sigma_{\mathrm{rand}} = 2.5$ around $t/\tau_{L0} = 45.2$.

To rationalize the observations of particle acceleration and clustering in decaying flow, we nondimensionalize Eq.~\ref{eq:StParticles} using the evolving Kolmogorov scales, $\eta(t)$ and $\tau_\eta(t)$, in  differential form. The decaying (non-stationary) flow necessitates using a differential non-dimensionalization, given by {$d\tilde{\bm{x}} = d\bm{x}/\eta(t)$ and $d\tilde{t} = dt/\tau_{\eta}(t)$}, such that the time and space increments co-evolve with the turbulence scales, analogous to the conformal time coordinate employed in cosmology to follow an expanding background. With $\tilde{t} = \int_{0}^{t} \frac{dt'}{\tau_{\eta}(t')}$, this leads to a time-translation invariant form for the non-dimensional equation of motion:
\vspace{-1pt}
\begin{equation}
\begin{aligned}
\frac{d \tilde{\bm{v}}}{d\tilde{t}} =   \beta  \frac{D \tilde{\bm{u}}}{D \tilde{t}}  + \frac{1}{\text{St}_{\eta e}(t)}( \tilde{\bm{u}} -  \tilde{\bm{v}} ) -  \frac 12 \dot{\tau}_{\eta}(t)  \left( \beta \tilde{\bm{u}}  - \tilde{\bm{v}} \right),
    \label{eq:NondimParticle}
\end{aligned}
\end{equation}
where $\beta = 3 \rho_f/(2 \rho_p + \rho_f)$ is the effective particle density ratio, and $\mathrm{St}_{\eta e}(t) \equiv r_p^2/(3\beta\nu \tau_\eta(t))$ is the instantaneous Kolmogorov-scale Stokes number
(see Sec. \ref{sec:NondimMaxey}). {Crucially, this choice of non-dimensionalization gives two quasi-stationary  terms that are analogous to those of SST, and an additional unsteady term, $\frac 12 \dot{\tau}_{\eta}(t)  \left( \beta \tilde{\bm{u}}  - \tilde{\bm{v}} \right)$, that evolves monotonically in time.}

 {For FDT, the dissipative time scale $\tau_\eta(t)=(\nu/\varepsilon(t))^{1/2}$ evolves analogously to Eq.~\ref{eq:KolmEvol}, $\tau_\eta(t)/\tau_{\eta 0} = 1 + C_\eta\,(t/\tau_{L0})^{\alpha}$ (Fig.~\ref{fig:flowDecay}).
Using this time-scale parameterization, an order-of-magnitude comparison of the terms in Eq.~\ref{eq:NondimParticle} indicates a critical time, {$t_{crit}$, such that Re$_\lambda (t_{crit})\sim \mathcal O(1)$} (see Supplemental Material, Sec.~\ref{sec:NondimMaxey}--\ref{sec:tcrit}). {For $t \to t_{crit}$, the evolution rate term becomes comparable to the turbulent forcing time scale.}  {For $t\ll t_{crit}$, the evolution-rate term is negligible and Eq.~\ref{eq:NondimParticle} reduces to its SST form, with the instantaneous Stokes number $\mathrm{St}_{\eta e}(t)$. The dynamic rescaling therefore maps the particle acceleration response onto the corresponding SST response.} For the present experiments and DNS, the $\dot{\tau}_\eta(t)$ term in Eq.~\ref{eq:NondimParticle} remains negligible throughout the decay phase. 

To model the temporal {decay} of the bubble and particle acceleration variances (Fig.~\ref{fig:collection}a), {we consider the flow sampled by the particles as monochromatic at the evolving Kolmogorov frequency $1/\tau_\eta(t)$ (Sec.~\ref{sec:SingleParticleAccVar} of the Supplemental Material \cite{supp}). This leads to a prediction for the evolving normalized acceleration variance:}

\begin{equation}
\gamma(\beta,\text{St}_{\eta e}(t)) \equiv \frac{\left < a^2_p \right >}{\left<a_{fp}^2\right>}   = \frac{\beta^2\text{St}_{\eta e}^2(t)  + 1}{\text{St}_{\eta e}^2(t)  + 1}.
\label{eq:response_accel}
\end{equation}

As shown in Fig.~\ref{fig:3}a, the predictions {(purple and dark-gray curves)} reasonably agree with the observations for both bubbles (DNS) and heavy particles (EXP) spanning $\beta \in [0,3] $ and $\text{St}_{\eta 0} \in [0.1, 25]$. {More complex response functions for the acceleration variance were obtained for SST, containing Reynolds number corrections \cite{zaichik_model_2011,zhang2019ModelDynamicsMicrobubblesa}.}  %
However, the deduction that matters here, by employing a monochromatic model (Eq.~\ref{eq:response_accel}), is that the evolving Stokes number $\mathrm{St}_{\eta e}(t)$ collapses the decaying data, establishing a quasi-stationary mapping of the acceleration statistics onto their SST counterparts. Small deviations are noticed at late times, in Fig.~\ref{fig:3}a, for both bubbles (DNS) and heavy particles (EXP). These may indicate the $\dot{\tau}_\eta(t)$ term of Eq.~\ref{eq:NondimParticle} becoming appreciable as $t\to t_{crit}$ (Sec.~\ref{sec:tcrit}). Note that in Fig.~\ref{fig:3}a, the inverse Stokes number was chosen for the horizontal axis because $\text{St}_{\eta e}^{-1}(t)$ grows during the decay; hence it serves as a proxy for time.

 {{Next, we test the validity of the dynamic rescaling for the clustering.}}
We distinguish $D_2$, the exponent of the pair correlation function at vanishing separation, $g(r)\sim r^{D_2-3}$, from a scale-specific correlation dimension $d_2(\ell)$ \cite{bec2008StochasticSuspensionsHeavy}, its local approximation at a finite separation $\ell$ {(End Matter, Appendix~D)}. Figure~\ref{fig:3}b,c shows the evolution of $d_2|_{\ell_c}(t)$, evaluated on a quasi-2D slice at a fixed separation scale $\ell_c = 10 \eta_0$, for heavy particles ($\beta \lesssim 0.01$) from EXP and DNS, respectively. Across the EXP and DNS datasets, the strongest clustering (minima of $d_2(t)$) is seen to coincide with $\text{St}_{\eta e}(t) \approx 1$. Thus, perhaps counterintuitively, decay of turbulence can give rise to a clustering intensification.

 As shown in Fig.~\ref{fig:3}b,c (see also Fig.~\ref{fig:D2_beta3_noncollapse} in the Supplemental Material \cite{supp}), when the correlation dimension is evaluated at a fixed length scale $\ell_c$ in FDT, the data does not fully collapse onto the SST response.  However, within our dynamic rescaling framework (Eq.~\ref{eq:NondimParticle}), the natural variables are the dimensionless separation, $\tilde{\ell}\equiv\ell/\eta(t)$, and the evolving Stokes number, $\text{St}_{\eta e}(t)$. This implies that
\begin{equation}
d_2(\tilde{\ell},t, \beta) \equiv d_2^{\rm SST}\!\left(\tilde{\ell};\,\mathrm{St}_{\eta e}(t),\beta\right),
\label{eq:d2_QS}
\end{equation}
i.e., the clustering in FDT measured for a co-evolving scale, $\tilde{\ell}$, and $\mathrm{St}_{\eta e}(t)$ can be mapped onto its SST counterpart. We consider a co-evolving separation scale $\tilde{\ell} = \ell/\eta(t)\approx 10$. Figure~\ref{fig:4new} shows that $d_2(\tilde{\ell},t, \beta)$ computed for bubbles (DNS) collapses when plotted against $\mathrm{St}_{\eta e}^{-1}(t)$, with the SST data (purple crosses) overlying the decay (FDT) data. Remarkably, these collapses extend to  the full range of $\beta$, spanning bubbles, light particles, and heavy particles, and covering two orders of magnitude in $\text{St}_{\eta 0} \in [0.1, 10]$ (see Fig.~\ref{fig:D2All_Re180_mov}a,b). The small Stokes limit, where we expect $(3 - d_2)  = \kappa \langle(\nabla \cdot \boldsymbol v)^2\rangle$ \cite{durham_turbulence_2013,supp}, yields a prediction  $(3-d_2) \sim {\text{St}_{\eta e}^{2}(t)}$ in agreement with the inset to Fig.~\ref{fig:4new}.

\begin{figure}[!tbp]
\centering
\includegraphics[width = 0.48\textwidth]{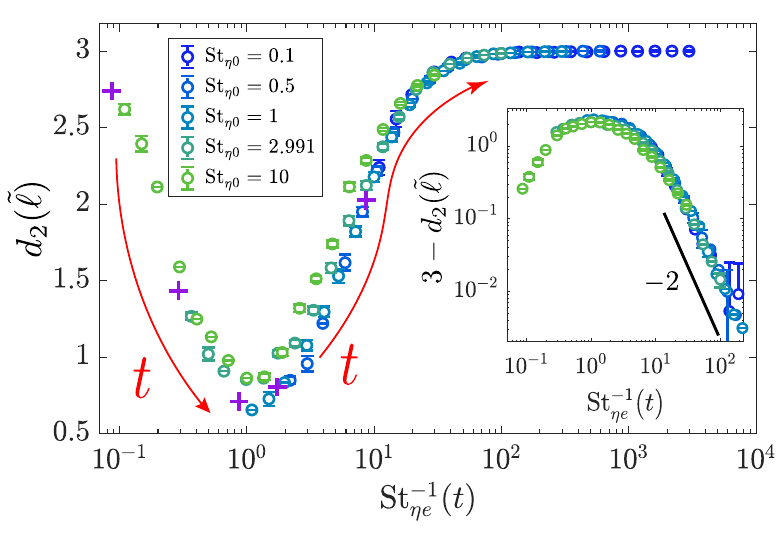}%
\caption{Dynamic rescaling collapses the clustering response across particle inertia and decay stages: $d_2(\tilde{\ell})$ vs.\ $1/\text{St}_{\eta e}(t)$ for bubbles ($\beta=3$), at a co-evolving scale $\ell_c(t)\approx 10\,\eta(t)$ ($\tilde{\ell}\approx10$), $\text{St}_{\eta_0}\in[0.1,10]$; the full $\beta$ range is shown in Fig.~\ref{fig:D2All_Re180_mov}. Purple crosses: SST data overlying the decay data. Inset: $3-d_2$ vs.\ $\text{St}_{\eta e}^{-1}(t)$ (log--log), with the predicted $-2$ slope.}
 \label{fig:4new}
  \vspace{-.4cm}
 \end{figure}
 
\ZY{In summary, we have employed a combination of laboratory experiments and DNS of decaying isotropic turbulence to reveal how inertial bubbles and particles evolve both individually and collectively during} the flow decay. The single-particle statistics adjust quasi-instantaneously to  the evolving $\tau_\eta(t)$, with only a weak dependence on the rate of turbulence decay, $\dot{\tau}_{\eta}(t)$.
 The clustering response can also be captured from a quasi-stationary framework, provided that the separation at which it is measured is dynamically rescaled with the evolving Kolmogorov scales. Thus, the scale-specific $d_2(\ell)$, evaluated on a growing dissipative scale $\ell(t) = C \eta(t)$, gives a full collapse of the data for bubbles, light particles and heavy particles,  spanning nearly two orders of magnitude in particle inertia. We show that memory effects, in both accelerations and clustering, become prominent only when the particle response time approaches the evolving turnover time of the flow (Eq.~\ref{eq:taupcrit}). {The non-monotonic intensification of clustering during decay reflects the passage of the evolving Stokes number through unity. Such a transient enhancement of clustering may have consequences for collision and contact rates of droplets in atmospheric clouds.}
}\\

\noindent
 {\color{black}
 DLN, ZY and LY contributed equally to this work and are co-first authors.
We thank Martin Maxey, S. Balachandar and  Olivier Simonin for fruitful discussions. We are grateful to Yinxiang Hu and Siyi Zhang for help during the experiments. VM acknowledges support from NSF Grant 2340593.

 }
	
\bibliographystyle{apsrev4-2}
\bibliography{references}

\clearpage
\onecolumngrid
\begin{center}
{\textbf{END MATTER}}
\end{center}
\vspace{4pt}
\twocolumngrid

\section*{Appendix A: Experimental details}
The turbulence chamber has eight motors positioned on the octahedral faces of the cubeoctahedron, each operated at a rotational speed of 6000 rpm. The central measurement volume is around $70 \times 70 \times 70$ mm$^3$, where the flow is nearly isotropic, with $u_y'/u_x' = 1.04$ and $u_z'/u_x' = 1.05$. The apparatus was designed to allow seeding of the flow with droplets, tracer particles, and  heavy (inertial) particles. Three to five ensembles of the decay were run to obtain converged statistics. In these experiments, the decay phase occurs when the gravitational settling effects are not dominant, as identified by evaluating settling number, $Sv(t) \equiv u'(t)/v_t > 1$, where $v_t$ is the terminal velocity of the particle. Furthermore, our observations indicated that the clusters do not have an appreciable structure bias in the $g$ direction. Gravitational settling effects  \cite{bec2014gravity,sarkar2025preferential} could become important in the very late stages of turbulence decay. This will be part of a future study. 

\section*{Appendix B: Numerical simulation details}
\setcounter{equation}{0}\renewcommand{\theequation}{B\arabic{equation}}
The DNS solve the incompressible Navier--Stokes equations
\begin{equation}
   \partial_t \boldsymbol{u} + \boldsymbol{u}\cdot \nabla \boldsymbol{u} = -\rho^{-1}\nabla p + \nu \nabla^2 \boldsymbol{u} + \boldsymbol{f}, \quad \nabla \cdot \boldsymbol{u} = 0,
   \label{eq:NS}
\end{equation}
where $\boldsymbol{u}$ is the flow velocity, $t$ is time, $p$ is the fluid pressure, $\rho$ is the fluid mass density, $\nu$ is the kinematic viscosity, and $\boldsymbol{f}$ a random forcing (per unit mass) acting on large scales as in Eswaran and Pope \cite{eswaran1988examination}. The forcing was set with the same amplitude and time-decorrelation as Ref.~\cite{chouippe_forcing_2015}. Here, $u'=\sqrt{\langle u_i^2\rangle/3}$ is the single-component root-mean-square ({\it rms}) flow velocity and $\lambda \equiv \sqrt{{15 u'^2 \nu}/{\varepsilon}}$ is the Taylor microscale, with $\varepsilon$ the energy dissipation rate. The particle density ratio was varied over the broadest possible range, from bubbles ($\rho_p/\rho_f \to 0$) to the aerosol limit ($\rho_p/\rho_f \gg 1$). Similarly, the particle size was varied over $r_p^* = [5 \times 10^{-3} - 2.9 \times 10^{-2}]$, where $r_p^*\equiv r_p/L$ is the particle radius normalized by the integral length scale. For each particle parameter combination [$\rho_p/\rho_f, r_p^*$], a total of $6.4\times 10^4$ particles were injected into the flow, allowed to reach statistical stationarity, and then evolved under decaying conditions. Further details of the numerical setup are given in Sec.~\ref{sec:NumericalDetails} of the Supplemental Material \cite{supp}.

\section*{Appendix C: Magnitude of the flow decay}
With the turbulent kinetic energy, $dk(t)/dt = -\varepsilon(t)$, the decay exponents follow as $\varepsilon(t) \sim t^{-3n/2}$ and $u'(t) \sim t^{-n/2}$. Consider ten integral times ($t/\tau_{L0}=10$) after the forcing is stopped: Eq.~\ref{eq:KolmEvol} gives $\eta/\eta_{0}\approx 4$, so the dissipative scale has expanded four-fold and the inertial-range scale separation $L/\eta$ has contracted by the same factor toward unity. Equivalently, since $\varepsilon\propto\eta^{-4}$, the dissipation rate and turbulent kinetic energy have dropped significantly ($\varepsilon/\varepsilon_{0}\sim 10^{-2}$ and $k/k_{0}\sim 10^{-2}$), the rms vorticity ($\omega_{\rm rms}\propto\sqrt{\varepsilon/\nu}$) is reduced by a factor of 10, and the rms fluid acceleration ($a_{\rm rms}\propto\varepsilon^{3/4}\nu^{-1/4}$) by a factor of 60. The Taylor Reynolds number $\mathrm{Re}_{\lambda}(t)\sim\mathrm{Re}_{\lambda 0}\,(t/\tau_{L0})^{-1/2}$ falls from $\mathrm{Re}_{\lambda 0} = 470$ to 150 in the experiments, and from $\mathrm{Re}_{\lambda 0}=180$ to 65 in the DNS. These monotonically decaying flow quantities can also be appreciated from the two insets to Fig.~\ref{fig-1}e corresponding to $t/\tau_{L0} \sim 10$.

\section*{Appendix D: Correlation dimension estimation}
\setcounter{equation}{0}\renewcommand{\theequation}{D\arabic{equation}}
The scale-specific correlation dimension at separation $\ell$ is estimated as
\begin{equation}
d_2(\ell)\equiv \frac{\ell^3g(\ell)}{\int_0^\ell r^2 g(r)\,\textrm{d}r},
\end{equation}
with $g(r)$ the pair correlation function and $\ell$ the separation at which $d_2$ is evaluated \cite{calzavarini_dimensionality_2008,bec2007HeavyParticleConcentration,bragg2015mechanisms}; note that $\lim_{\ell \to 0} d_2(\ell) = D_2$.

\onecolumngrid
\clearpage


\setcounter{equation}{0}
\setcounter{section}{0}
\setcounter{figure}{0}
\setcounter{table}{0}
\makeatletter
\renewcommand{\theequation}{S-\arabic{equation}}
\renewcommand{\thefigure}{S-\arabic{figure}}
\renewcommand{\thesection}{S-\Roman{section}}
\renewcommand{\thesubsection}{S-\Roman{section}-\alph{subsection}}
\renewcommand{\thetable}{S-\arabic{table}}
\preprint{AIP/123-QED}

\title{Supplemental  Material}
\maketitle

\clearpage
\newpage

\setcounter{equation}{0}
\setcounter{section}{0}
\setcounter{figure}{0}
\setcounter{table}{0}
\setcounter{page}{1}
\makeatletter
\renewcommand{\theequation}{S-\arabic{equation}}
\renewcommand{\thefigure}{S-\arabic{figure}}
\renewcommand{\thesection}{S-\roman{section}}
\renewcommand{\thesubsection}{S-\roman{section}-\alph{subsection}}
\renewcommand{\thetable}{S-\arabic{table}}
\preprint{AIP/123-QED}
\setcounter{secnumdepth}{3}

\onecolumngrid
\vspace*{1cm}
\begin{center}
    {\large {\bf SUPPLEMENTAL MATERIAL}}
\end{center}
\vspace{0.5cm}

\title{Supplemental  Material}
\maketitle

\section{Numerical details}
\label{sec:NumericalDetails}

\DLN{To create the initial state of the decay, a homogeneous isotropic turbulent flow is first simulated within a cubic box of side length $H$ with periodic boundaries. The incompressible Navier--Stokes equations,
\begin{flalign}
    \frac{\partial \boldsymbol{u}}{\partial t} + \boldsymbol{u}\cdot \nabla \boldsymbol{u} = -\frac{1}{\rho}\nabla p + \nu \nabla^2 \boldsymbol{u} + \boldsymbol{f}, \\
    \nabla \cdot \boldsymbol{u} = 0,
\end{flalign}
where $\boldsymbol{u}$ is the flow velocity, $t$ is time, $p$ is {the} fluid pressure, $\rho$ is the fluid mass density, $\nu$ is the kinematic viscosity, and $\boldsymbol{f}$ denotes the randomized external forcing \cite{eswaran1988examination}, 
are directly simulated in space using energy-conserving second-order centered finite differences integrated in time using a fractional-step third-order Runge--Kutta scheme for the nonlinear terms and an implicit Crank--Nicolson scheme for the viscous terms. The code is {open-sourced} under the name AFiD \cite{van_der_poel_pencil_2015}, and has been used for the simulation of triply periodic turbulence \cite{spandan2020fluctuation} among others. The computational grid resolution is chosen as a $512^3$ staggered grid, to ensure that the flow is well resolved as $\kappa_{\text{max}}\eta > 2$, where $\kappa_{\text{max}}$ is the maximum wave number in the flow in a single direction. The time step is dynamically chosen so that the maximum Courant--Friedrichs--Lewy condition number is 1.2.}

 \DLN{The simulation is started from zero initial conditions. Upon start of the simulation, the flow is forced through the large-scale force vector $\boldsymbol{f}$, which forces all modes whose wave number $\kappa$ is smaller than a chosen $\kappa_f$. In practice this is taken as $\kappa_f / \kappa_0 = 2.3$, where $\kappa_0 = 2\pi/H$ is the largest possible wave number in any direction. The direction and magnitude of this forcing is calculated based on random processes which drive the time evolution of these selected modes based on a target energy flux $\epsilon^*$ and force correlation time $T_L$. Depending on the choice of these two parameters, one obtains the desired initial Taylor-Reynolds number Re$_{\lambda0}\equiv \frac{u'\lambda}{\nu}$, with $u'$ the root-mean-square flow velocity and $\lambda \equiv \sqrt{\frac{15 \nu}{\varepsilon}}u'$ the Taylor microscale length with $\varepsilon$ the flow's total energy dissipation rate (see \cite{eswaran1988examination} for more details). We choose the same parameters as Refs.~\cite{chouippe_forcing_2015,spandan2020fluctuation} to achieve the desired Re$_{\lambda0}$. The forcing is applied until a statistically stationary state is reached, or for about four initial large-eddy time scales $\tau_{L0}$. To simulate the decay, the forcing is then turned off, and data are collected until Re$_\lambda \approx 13$, which was observed to be sufficient to capture the full decay process.}  The decay of $\varepsilon(t)$ and $u'$ {is} shown in Fig.~\ref{fig:flowDecay}. The inset shows that both follow robust power laws, with $u'^2 \sim t^{-n}$ and $\varepsilon \sim t^{-3n/2}$. The integral length scale $L$ can be given by $L=C_\varepsilon u'^3(t)/\varepsilon(t)$, where $C_\varepsilon$ is often taken as a constant \cite{panickacheril_john_laws_2022}.
\begin{figure}[!hbp]
\centering
\includegraphics[width = 3.39in]{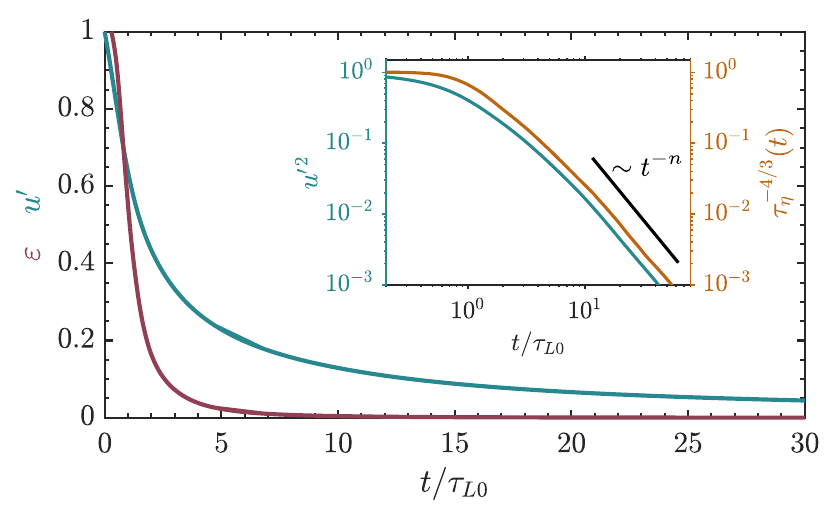} 
 \caption{\DLN{Decay of velocity fluctuations $u'(t)$ and $\varepsilon(t)$ over large-eddy-normalized time $t/\tau_{L0}$, Re$_{\lambda0}=180$. Inset shows $u'(t)$ and $\tau_\eta(t)$ decay with log scale. All the quantities shown in this figure are normalized with their initial values $u'_0$, $\varepsilon_0$, and $\tau_{\eta0}$.}}
 \label{fig:flowDecay}
 \end{figure}

\section{Nondimensionalization of Maxey--Riley equation in decaying turbulence}
\label{sec:NondimMaxey}
\providecommand{\del}[1]{\textcolor{olive}{\sout{#1}}}
\providecommand{\new}[1]{{\color{violet}#1}}
\providecolor{vfivecolor}{rgb}{0.80,0.33,0.0}
\providecommand{\vfive}[1]{\textcolor{vfivecolor}{#1}}
\providecolor{v5light}{rgb}{0.93,0.62,0.38}
\providecommand{\vfcut}[1]{\textcolor{v5light}{\sout{#1}}}
\providecommand{\vfix}[1]{\textcolor{red}{\ifmmode\bm{#1}\else\textbf{#1}\fi}}
\providecommand{\edit}[1]{\textcolor{red}{#1}}
\providecolor{vfinal}{rgb}{0.55,0.0,0.85}
\providecolor{vfinallight}{rgb}{0.80,0.62,0.92}
\providecommand{\pfix}[1]{\textcolor{vfinal}{#1}}
\providecommand{\pcut}[1]{\textcolor{vfinallight}{\sout{#1}}}

We nondimensionalize the point-particle equation of motion, Eq.~\ref{eq:StParticles}, with the evolving Kolmogorov scales in differential form, keeping only $\bm x$, $t$, $\bm u$, and $\bm v$:
\begin{equation}
    \tilde{\bm u}=\frac{\bm u}{u_\eta(t)},\qquad
    \tilde{\bm v}=\frac{\bm v}{u_\eta(t)},\qquad
    d\tilde{\bm x}=\frac{d\bm x}{\eta(t)},\qquad
    d\tilde t=\frac{d t}{\tau_\eta(t)}.
    \label{eq:S3defs}
\end{equation}
The differential form is chosen so that the nondimensional velocity keeps $\tilde{\bm v}\approx d\tilde{\bm x}/d\tilde t$, as shown below.

Dividing Eq.~\ref{eq:StParticles} by the instantaneous Kolmogorov acceleration $a_\eta(t)=(\varepsilon^3/\nu)^{1/4}=u_\eta/\tau_\eta$ puts the particle acceleration reported in the main text on the left-hand side,
\begin{equation}
    \frac{d\bm v}{d t}\Big/a_\eta
    =\beta\frac{D\bm u}{D t}\Big/a_\eta
    +\frac{1}{\text{St}_{\eta e}(t)}\left(\bm u-\bm v\right)\Big/a_\eta .
    \label{eq:S4plotted}
\end{equation}
These terms can be measured directly, and the balance already resembles the stationary one. Yet the resemblance is not sufficient, because the left-hand side is $(d\bm v/d t)/a_\eta$ rather than the derivative of the nondimensional velocity $d\tilde{\bm v}/d\tilde t$; and it is also similar for the $(d\bm u/d t)/a_\eta$ term. Eq.~\ref{eq:S4plotted} is then not closed and cannot be integrated for $\bm v(t)$ without $a_\eta(t)$. We shall ask when the terms in the equation (``measured'' quantities) and nondimensional velocity derivatives coincide. Using $\dot u_\eta/u_\eta=-\frac12\dot\tau_\eta/\tau_\eta$, we have
\begin{equation}
    \frac{d\bm v}{d t}\Big/a_\eta=\frac{d\tilde{\bm v}}{d\tilde t}-\frac12\dot\tau_\eta(t)\,\tilde{\bm v},
    \qquad
    \beta\frac{D\bm u}{D t}\Big/a_\eta=\beta\frac{D\tilde{\bm u}}{D\tilde t}-\frac12\beta\dot\tau_\eta(t)\,\tilde{\bm u}.
    \label{eq:S5map}
\end{equation}
The $\dot\tau_\eta$ terms are negligible when $d\tilde{\bm v}/d\tilde t\gg\dot\tau_\eta\tilde{\bm v}$ and $D\tilde{\bm u}/D\tilde t\gg\dot\tau_\eta\tilde{\bm u}$. Since $\bm v\sim\bm u$ and $d\tilde{\bm v}/d\tilde t,\ D\tilde{\bm u}/D\tilde t\sim\tilde a_\eta(t)\equiv1$, this corresponds to
\begin{equation}
    \dot\tau_\eta(t)\,\tilde{\bm u}\ll1,
    \label{eq:negcond}
\end{equation}
which reduces to (proof in Sec.~\ref{sec:tcrit})
\begin{equation}
    t\ll t_{crit},\qquad\text{where}\ \ \text{Re}_\lambda(t_{crit})\sim\mathcal{O}(1).
    \label{eq:tcritcond}
\end{equation}
For $t\ll t_{crit}$ the measured quantities agree with the nondimensional derivatives. \\

In the general case, the equation can be rewritten in the closed, solvable form
\begin{equation}
    \frac{d\tilde{\bm v}}{d\tilde t}
    =\underbrace{\beta\frac{D\tilde{\bm u}}{D\tilde t}}_{\bm a_{fluid}}
    +\underbrace{\frac{1}{\text{St}_{\eta e}(t)}(\tilde{\bm u}-\tilde{\bm v})}_{\bm a_{drag}}
    -\underbrace{\frac12\dot\tau_\eta(t)\bigl(\beta\tilde{\bm u}-\tilde{\bm v}\bigr)}_{\bm a_{non-QS}} ,
    \label{eq:nondimMR}
\end{equation}
with $\bm a_{fluid}$ the quasi-stationary fluid parcel acceleration, $\bm a_{drag}$ the drag, and $\bm a_{non-QS}$ the correction carrying $\dot\tau_\eta$. All three are solvable from Eq.~\ref{eq:nondimMR} and, for $t\ll t_{crit}$, stay within negligible deviation of the measured quantities.

We now ask when $\bm a_{non-QS}$ can be dropped from Eq.~\ref{eq:nondimMR}, so that the equation takes the stationary form
\begin{equation}
    \frac{d\tilde{\bm v}}{d\tilde t}\approx\beta\frac{D\tilde{\bm u}}{D\tilde t}+\frac{1}{\text{St}_{\eta e}(t)}(\tilde{\bm u}-\tilde{\bm v}),
    \label{eq:sssform}
\end{equation}
the statistically stationary equation evaluated at the instantaneous Stokes number $\text{St}_{\eta e}(t)$. This depends on the size of $\bm a_{non-QS}$ relative to $\bm a_{fluid}+\bm a_{drag}$, which we estimate in three regimes.

(a) $\text{St}_{\eta e}\ll1$. The first-order perturbation solution of the Maxey--Riley equation, Eq.~\ref{eq:sssform}, is given by 
\begin{equation}
    \tilde{\bm v}=\tilde{\bm u}+\text{St}_{\eta e}(\beta-1)\frac{D\tilde{\bm u}}{Dt}+O(\text{St}_{\eta e}^2),
    \label{eq:pert_a}
\end{equation}
which we will discuss in Sec.~\ref{sec:Perturb}. This gives
\begin{equation}
    \tilde{\bm u}-\tilde{\bm v}\approx\text{St}_{\eta e}{(1-\beta)}\frac{D\tilde{\bm u}}{Dt}\sim\text{St}_{\eta e}{(1-\beta)},
    \label{eq:uv_a}
\end{equation}
so
\begin{equation}
    \bm a_{fluid}\sim\beta,\qquad
    \bm a_{drag}\sim(\beta-1),\qquad
    \bm a_{non-QS}\sim\dot\tau_\eta(t)\,\tilde{\bm u}.
    \label{eq:scal_a}
\end{equation}
The first two sum to order unity, so $\bm a_{non-QS}$ is negligible when $\dot\tau_\eta(t)\tilde{\bm u}\ll1$, which is Eq.~\ref{eq:negcond}, satisfied also for $t\ll t_{crit}$.

(b) $\text{St}_{\eta e}\gtrsim1$, finite $\beta$. Here $\tilde{\bm u}-\tilde{\bm v}\sim\tilde u'(t)$, so the drag term is
\begin{equation}
    \bm a_{drag}\sim\frac{\tilde u'}{\text{St}_{\eta e}}.
    \label{eq:drag_b}
\end{equation}
Since the Stokes number is large, $\bm a_{fluid}\gtrsim \bm a_{drag}$, and the comparison is again between $\bm a_{fluid}$ and $\bm a_{non-QS}$. This returns the same condition as case (a), $t\ll t_{crit}$.

(c) $\text{St}_{\eta e}\gtrsim1$, $\beta\to0$. The fluid term $\beta\bm a_{fluid}\to0$, so the comparison is between $\bm a_{non-QS}$ and $\bm a_{drag}$. With $\beta\tilde{\bm u}-\tilde{\bm v}\sim\tilde u'$ and $\tilde{\bm u}-\tilde{\bm v}\sim\tilde u'$,
\begin{equation}
    \bm a_{non-QS}\sim\frac12\dot\tau_\eta(t)\,\tilde u',\qquad
    \bm a_{drag}\sim\frac{\tilde u'}{\text{St}_{\eta e}},
    \label{eq:scal_c}
\end{equation}
so $\bm a_{non-QS}<\bm a_{drag}$ requires $\tfrac12\dot\tau_\eta(t)\,\text{St}_{\eta e}<1$. With $\text{St}_{\eta e}=\tau_p/\tau_\eta(t)$ and $\text{St}_{\eta0}=\tau_p/\tau_{\eta0}$, this reads
\begin{equation}
    \frac{2\,\tau_\eta(t)}{\tau_{\eta0}\,\dot\tau_\eta(t)}>\text{St}_{\eta0}.
    \label{eq:heavycond}
\end{equation}
Using the decay fit $\tau_\eta(t)/\tau_{\eta0}=1+C_\eta(t/\tau_{L0})^\alpha$ with 
\begin{equation}
    \dot\tau_\eta(t)=C_\eta\alpha\left(\frac{t}{\tau_{L0}}\right)^{\alpha-1}\frac{\tau_{\eta0}}{\tau_{L0}},
    \label{eq:tauetadot}
\end{equation}
and $\tau_p=\text{St}_{\eta0}\tau_{\eta0}$, Eq.~\ref{eq:heavycond} becomes
\begin{equation}
    t+\frac{\tau_{L0}^{\,\alpha}}{C_\eta}\,t^{\,1-\alpha}>\frac{\alpha}{2}\,\tau_p .
    \label{eq:heavymid}
\end{equation}
Because $\alpha>1$, the left-hand side has a minimum in $t$,
\begin{equation}
    \mathrm{LHS}_{\min}=\frac{\alpha}{\alpha-1}\left(\frac{\alpha-1}{C_\eta}\right)^{1/\alpha}\tau_{L0},
    \label{eq:lhsmin}
\end{equation}
so the condition holds for all $t$ only if
\begin{equation}
    \boxed{\ \tau_p<\frac{2}{\alpha-1}\left(\frac{\alpha-1}{C_\eta}\right)^{1/\alpha}\tau_{L0}\sim\tau_{L0}.\ }
    \label{eq:taupcrit}
\end{equation}
Thus, for a large-St, heavy particle, the quasi-stationary reduction holds as long as the response time stays well below the large-eddy time, $\tau_p\ll\tau_{L0}$.

In summary, we have shown that the Maxey--Riley equation can be nondimensionalized in {differential form} using  the evolving Kolmogorov scales, and will reduce to the quasi-stationary form Eq.~\ref{eq:sssform} unless one of the two conditions listed below is broken. The first is late time, $t\gtrsim t_{crit}$, which is when $\text{Re}_\lambda(t)\sim1$ and the inertial range separating the Kolmogorov and large-eddy scales has closed. The particle acceleration is governed by the fast Kolmogorov-scale motions, whereas the decay is set by the slow large scales, so the particle feels the decay only once $\tau_\eta\sim\tau_L$. The second is large inertia, $\tau_p\gtrsim\tau_{L0}$ from Eq.~\ref{eq:taupcrit} for heavy particles. When the particle response time approaches the large-eddy time, the particle and flow statistics evolve at comparable rates, and history effects appear.

\section{Perturbative solution of the Maxey--Riley equation}\label{sec:Perturb}

When $\text{St}_\eta\ll 1$, particle velocity $\tilde{\bm{v}}$ can be seen as {a} single-valued function $\tilde{\bm{v}} = \tilde{\bm{v}}(\tilde{\bm{x}}, \tilde t)$ with $\left|\tilde{\bm{v}}-\tilde{\bm{u}}\right|\ll\left|\tilde{\bm{u}}\right|$. We will prove this below. Consider the relative velocity of particle to fluid $\tilde{\bm{w}} = \tilde{\bm{v}} -\tilde{\bm{u}}$. Notice that the derivative $d/dt$ refers to the derivative along the particle trajectory in the flow, giving

\begin{equation}
\begin{split}
    \frac{d\tilde{\bm{v}}}{dt} &= \frac{d\tilde{\bm{w}}}{dt} + \frac{d\tilde{\bm{u}}}{dt} \\
    &= \frac{d\tilde{\bm{w}}}{dt} + \frac{\partial\tilde{\bm{u}}}{\partial t} + (\tilde{\bm{v}}\cdot\nabla)\tilde{\bm{u}}\\
    &= \frac{d\tilde{\bm{w}}}{dt} + \frac{D\tilde{\bm{u}}}{Dt} + (\tilde{\bm{w}}\cdot\nabla)\tilde{\bm{u}}\ .
\end{split}
\end{equation}
Assuming Eq.~\ref{eq:sssform} is applicable leads to:

\begin{equation}\label{eq:maxeyriley3}
\begin{split}
    \frac{d\tilde{\bm{w}}}{dt} + \frac{\tilde{\bm{w}}}{\text{St}_{\eta}} &= (\beta - 1) \frac{D\tilde{\bm{u}}}{Dt} - (\tilde{\bm{w}} \cdot \nabla)\tilde{\bm{u}}. \\
\end{split}
\end{equation}

We can do perturbative expansion as $\tilde{\bm{w}} = \text{St}_{\eta} \tilde{\bm{w}}^{(1)}+ \text{St}_{\eta}^2 \tilde{\bm{w}}^{(2)}+...$, obtaining:

\begin{equation}
\begin{split}
  \text{St}_{\eta}^2\frac{d\tilde{\bm{w}}^{(1)}}{dt} + \text{St}_{\eta}\tilde{\bm{w}}^{(1)} 
  =& \text{St}_{\eta}(\beta - 1) \frac{D\tilde{\bm{u}}}{Dt} \\
  &\hspace{0.1cm}- \text{St}_{\eta}^2(\tilde{\bm{w}}^{(1)} \cdot \nabla)\tilde{\bm{u}}
  +O(\text{St}_{\eta}^3).
\end{split}
\end{equation}
Terms on the R.H.S., except the first one, are of order $\text{St}_{\eta}^2$ and can be ignored. For a statistically stationary system, we may also ignore the first term on the L.H.S. This gives
\begin{equation}
  \tilde{\bm{w}}^{(1)}= (\beta-1)\frac {D\tilde{\bm{u}}}{Dt},
\end{equation}
\begin{equation}
  \tilde{\bm{w}}= \text{St}_{\eta}(\beta-1)\frac {D\tilde{\bm{u}}}{Dt}+O(\text{St}_{\eta}^2).
\end{equation}
This satisfies our assumption $\left|\tilde{\bm{w}}\right|\ll\left|\tilde{\bm{u}}\right|$. 

\section{Criterion for the quasi-stationary reduction}
\label{sec:tcrit}

The quasi-stationary reduction of Sec.~\ref{sec:NondimMaxey} rests on the condition Eq.~\ref{eq:negcond}, $\dot\tau_\eta(t)\,\tilde{\bm u}\ll1$. We now turn it into a time criterion $t\ll t_{crit}$. Using Eq.~\ref{eq:tauetadot} for $\dot\tau_\eta$, together with
\begin{equation}
    \frac{\tau_{\eta0}}{\tau_{L0}}\sim\text{Re}_{\lambda0}^{-1},\qquad
    \tilde{\bm u}=\frac{\bm u}{u_\eta(t)}\sim\text{Re}_\lambda^{1/2}(t),
    \label{eq:tcrit_inputs}
\end{equation}
the condition Eq.~\ref{eq:negcond} reads
\begin{equation}
    C_\eta\alpha\left(\frac{t}{\tau_{L0}}\right)^{\alpha-1}\text{Re}_{\lambda0}^{-1}\text{Re}_\lambda^{1/2}(t)\ll1.
    \label{eq:tcrit_mid1}
\end{equation}
Using
\begin{equation}
    \text{Re}_\lambda\sim\frac{u'^2}{(\nu\epsilon)^{1/2}}=\frac{u'^2\,\tau_\eta}{\nu},
    \qquad
    \frac{\text{Re}_\lambda(t)}{\text{Re}_{\lambda 0}}=\frac{u'^2(t)}{u_0'^2}\frac{\tau_\eta(t)}{\tau_{\eta 0}},
\end{equation}
with
\begin{equation}
\left(\frac{\tau_\eta(t)}{\tau_{\eta0}}\right)^{2}
=\frac{\varepsilon_0}{\varepsilon(t)}
=\frac{\mathrm{d}u'^2/\mathrm{d}t\big|_{0}}{\mathrm{d}u'(t)^2/\mathrm{d}t},
\end{equation}
we have
\begin{equation}
    \frac{u'^2(t)}{u_0'^2}\approx C_u\Bigl(\frac{t}{\tau_{L0}}\Bigr)^{-2\alpha+1},\qquad C_u\sim \mathcal O(1).
\end{equation}
The instantaneous Taylor-scale Reynolds number thus decays as
\begin{equation}
    \frac{\text{Re}_\lambda(t)}{\text{Re}_{\lambda0}}\approx\left(\frac{t}{\tau_{L0}}\right)^{1-\alpha}.
    \label{eq:tcrit_redecay}
\end{equation}
Substituting Eq.~\ref{eq:tcrit_redecay} into Eq.~\ref{eq:tcrit_mid1} collapses the condition to
\begin{equation}
    C_\eta\alpha\left(\frac{t}{\tau_{L0}}\right)^{(\alpha-1)/2}\text{Re}_{\lambda0}^{-1/2}\ll1,
    \label{eq:tcrit_mid2}
\end{equation}
that is
\begin{equation}
  \boxed{t\ll t_{crit}=C_{crit}^{2/(\alpha-1)}\,\text{Re}_{\lambda0}^{1/(\alpha-1)}\,\tau_{L0},}
    \label{eq:tcrit_expr}
\end{equation}
where $C_{crit}$ is a constant $\sim\mathcal O(1)$. The prefactor $C_{crit}^{2/(\alpha-1)}$ is strongly sensitive to $\alpha$ as $\alpha$ can be close to 1. So rather than read it off we evaluate the turbulent strength characterized by $\text{Re}_\lambda$ at $t_{crit}$. Inserting Eq.~\ref{eq:tcrit_expr} into Eq.~\ref{eq:tcrit_redecay} gives
\begin{equation}
    \boxed{\text{Re}_\lambda(t_{crit})\approx\text{Re}_{\lambda0}\left(\frac{t_{crit}}{\tau_{L0}}\right)^{1-\alpha}=C_{crit}^{-2}\sim\mathcal{O}(1).}
    \label{eq:reltcrit}
\end{equation}
Equation~\ref{eq:reltcrit} is the clean statement of the criterion: it carries no $\text{Re}_{\lambda0}$, so the breakdown is fixed by the instantaneous turbulent state rather than the initial condition, and a larger $\text{Re}_{\lambda0}$ only postpones $t_{crit}$ without moving the threshold.

\section{Single particle acceleration variance}
\label{sec:SingleParticleAccVar}

{The derivation below adopts a single-frequency (monochromatic) idealization of the fluid acceleration seen along the particle path; a fuller justification with the measured acceleration spectrum and preferential sampling is deferred to a subsequent paper.}

{To simplify, we approximate the flow velocity as a monochromatic, homogeneous, sine field,
\begin{equation}
    \tilde{\bm u}(\tilde t)=\sqrt 2U\sin(\omega\tilde t),
\end{equation}
where $U$ is the non-dimensionalized velocity magnitude and $\omega$ is the frequency. The flow acceleration at particle location is thus given by
\begin{equation}
    \tilde{\bm a}_{fp}(\tilde t)=\sqrt 2U\omega\cos(\omega\tilde t).
\end{equation}}

{The particle equations of motion (\ref{eq:sssform}) are then 
\begin{equation} \frac{d \tilde{\bm{v}}}{d\tilde{t}} = \beta \sqrt 2U\omega\cos{(\omega \tilde{t})} + \frac{1}{\tau_p}\left(\sqrt 2U\sin{(\omega \tilde{t})} - \tilde{\bm{v}}\right).\end{equation}
This can be solved using standard methods to obtain 
\begin{equation} \begin{split}
    \tilde{\bm{v}}(\tilde{t})   &= \frac{\sqrt 2U }{(\tau_p^2 \omega^2 + 1)}  \left[\sin(\omega \tilde{t}) (\beta\tau_p^2\omega^2+1) \right. \\ 
    &\hspace{3cm}+ \left.\cos(\omega \tilde{t})(\beta-1)\tau_p\omega \right]
\end{split}\end{equation}
so the particle acceleration is 
\begin{equation}\begin{split}
    a_p(\tilde{t})&=\frac{\sqrt 2U\omega }{(\tau_p^2 \omega^2 + 1)}  \left[\cos(\omega \tilde{t}) (\beta\tau_p^2\omega^2+1) \right. \\ 
    &\hspace{3cm}- \left.\sin(\omega \tilde{t})(\beta-1)\tau_p\omega \right],
\end{split}\end{equation}
where one can take the second moment over one period $T=2\pi/\omega$ to obtain
\begin{equation}\langle a_p^2\rangle = \frac{(U\omega)^2}{\text{St}_\omega^2 + 1} \left(\beta^2\text{St}_\omega^2  + 1\right),\end{equation}
where St$_{\omega}\equiv \tau_p \omega$. One can then normalize by the second moment of the particle-sampled fluid acceleration $\langle a_{fp}^2 \rangle = (U\omega)^2$, and taking $\omega=1/\tau_\eta(t)$ yields}
\begin{equation}
\frac{\left < a^2_p \right >}{\left<a_{fp}^2\right>} = \frac{\beta^2\text{St}_{\eta e}(t)^2  + 1}{\text{St}_{\eta e}(t)^2 + 1},
\end{equation}
which is a function of $\beta$ and $\text{St}_{\eta e}$. 

\section{Clustering at low Stokes number}
\label{sec:ClusteringLowSt}

The accumulation of the particles can be seen as following the model \cite{durham_turbulence_2013,gustavsson2015AnalysisCorrelationDimension,bec2003FractalClusteringInertiala}:
\begin{equation}
    D_2 = 3-\kappa\left<(\nabla \cdot \tilde{\bm{v}})^2\right>
    =3-\kappa\left<\vartheta^2\right>,
    \label{eq_D2}
\end{equation}
where $\kappa$ is a constant related to the flow, and $\nabla\cdot\bm v=\nabla\cdot\bm w=\vartheta$. We may then obtain
\begin{equation}
    3-D_2 = \kappa(\beta-1)^2\text{St}_{\eta}^2{\langle S_0^2\rangle}\sim \text{St}_{\eta}^2,
    \label{eq_D2-1}
\end{equation}
where $S_0=\nabla\cdot[(\bm u\cdot\nabla)\bm u]$. This $\text{St}_{\eta}^2$ dependence is seen in the inset of Fig.~\ref{fig:4new}.

\begin{figure}[!htbp]
    \centering
    \includegraphics[width=0.52\textwidth]{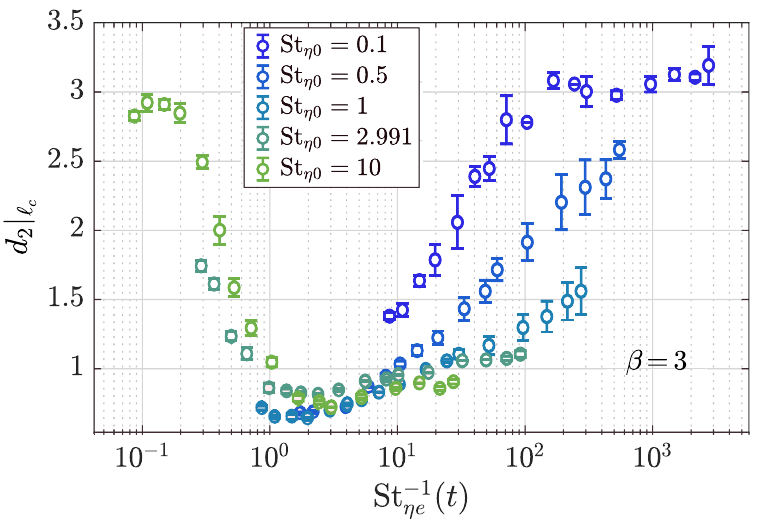}
    \caption{Correlation dimension at a fixed separation scale $\ell_c$, $d_2|_{\ell_c}$, vs.\ $\mathrm{St}_{\eta e}^{-1}(t)$ for bubbles ($\beta = 3$, DNS at $\mathrm{Re}_{\lambda 0} = 180$), spanning $\mathrm{St}_{\eta 0} \in [0.1, 10]$.}
    \label{fig:D2_beta3_noncollapse}
\end{figure}

\begin{figure}[!tbp]
    \centering
    \includegraphics[width= 0.75\textwidth]{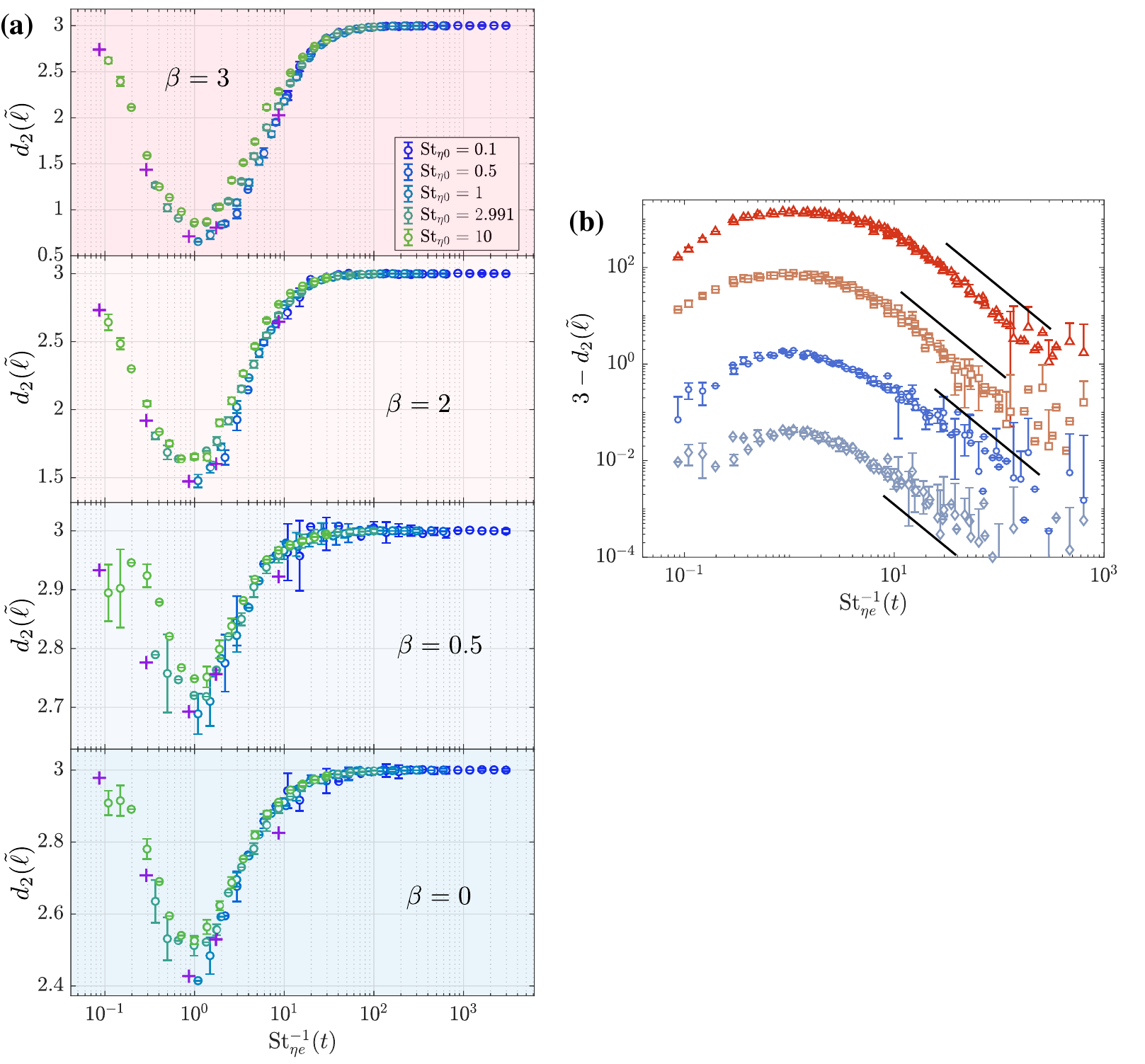}
    \caption{(a) $d_2$ data for all cases at Re$_{\lambda 0}=180$ at $\tilde{\ell} \equiv \ell/\eta(t) \approx 10$. The purple crosses indicate the start of each decay process. (b) $3-d_2$ data for all cases at Re$_{\lambda 0}=180$. The red, orange, blue, and gray {colors represent} $\beta =3, 2, 0,$ and $0.5$, respectively. For clarity, different $\beta$ series were vertically offset in (b).}
    \label{fig:D2All_Re180_mov}
\end{figure}

\section{$d_2$ maps for different particle densities}
\label{sec:D2DiffParticleDensities}

Fig.~\ref{fig:D2_beta3_noncollapse} shows the correlation dimension evaluated at a fixed separation scale, $d_2|_{\ell_c}$, plotted against $\mathrm{St}_{\eta e}^{-1}(t)$ for bubbles ($\beta = 3$). A direct mapping based on $\mathrm{St}_{\eta e}(t)$ fails to collapse the decay curves; a similar failure to collapse is seen for other particle densities (not shown here). This further highlights the need for a dynamic rescaling based on  differentially evolving time and spatial scales. The dynamically rescaled natural variable is the dimensionless separation $\tilde{\ell}\equiv\ell/\eta(t)$, and the QS mapping implies that\begin{equation}
d_2(\tilde\ell,t) \equiv d_2^{\text{SST}}\!\left(\tilde{\ell};\,\mathrm{St}_{\eta e}(t),\beta\right).
\end{equation}
Here, $\tilde{\ell}$ is an evolving scale that varies in time $\propto \eta(t)$. Fig.~\ref{fig:D2All_Re180_mov}a shows $d_2$ computed over an \emph{evolving} length scale $\tilde{\ell} \equiv \ell/\eta(t) \approx 10$. We exhaustively span the full range of particle densities and  two orders of magnitude in Stokes number.

\end{document}